\documentclass[jkps,preprint,fleqn,showpacs,showkeys]{revtex4}
\usepackage{graphicx}
\usepackage{subcaption}
\usepackage{amssymb}
\usepackage{amsmath}
\usepackage{bm}
\usepackage{kotex}
\usepackage{upgreek} 
\usepackage{libertine} 
\usepackage{siunitx} 
\usepackage{capt-of}
\usepackage[labelsep=period]{caption}

\begin{document}
\title{Neutron Detection Using a Gadolinium-Cathode GEM Detector}
\author{DongHyun \surname{Song}}
\email{donghyun.song@cern.ch}
\author{Kyungeon \surname{Choi}}
\author{Youngun \surname{Jeng}}
\author{Yechan~\surname{Kang}}
\author{Jason SangHun \surname{Lee}}
\author{Inkyu \surname{Park}}
\email{icpark@uos.ac.kr}
\author{Sunyoung \surname{Yoo}}
\affiliation{Department of Physics, University of Seoul, Seoul 02504, Korea}

\begin{abstract}
A gas electron multiplier (GEM) detector with a gadolinium cathode has been developed to explore its potential application as a neutron detector.
It consists of three standard-sized ($10\times 10$ cm${}^{2}$) GEM foils and a thin gadolinium plate as the cathode, which is used as a neutron converter.
The neutron detection efficiencies were measured for two different cathode setups and for two different drift gaps. 
The thermal neutron source at the Korea Research Institute of Standards and Science (KRISS) was used to measure the neutron detection efficiency.
Based on the neutron flux measured by KRISS, the neutron detection efficiency of our gadolinium GEM detector was $4.630 \pm 0.034(stat.) \pm 0.279(syst.) \%$.
\end{abstract}

\pacs{29.40.Cs , 87.53.Qc, 29.40.−n}
\keywords{GEM, Neutron, Gadolinium, Particle detector}

\maketitle

\begin{center}
\section{INTRODUCTION}
\end{center}

Neutron are of interest in nuclear and particle physics as well as in material science and have various applications such as non-destructive testing and neutron therapy \cite{Neutron theraphy, NeutronApplication_book, Non-destructive Test _ book}. 
As the demand for neutron sources is growing, accelerator-based spallation facilities are expanding in some countries, including the United States, European Countries and China aiming to increase the flux of neutron beams \cite{USA spallation neutron source, European spallation source, China spallation neutron source}.
Currently, boron neutron capture therapy (BNCT)~\cite{EORTC_BNCT, Clinical_BNCT} is gaining attention as the next-generation radiation therapy technology, and several neutron facilities are already available in Japan and are planned in Korea \cite{Japanese society BNCT, dawon_medacs}.
Therefore, the development of a low-cost neutron detector with high efficiency has become important.
As neutrons do not ionize, elements that react well with neutrons, such as helium, lithium, or boron, as mediators to detect neutrons must be used.
The most commonly used neutron detector is usually a gas detector filled with $^3$He or BF$_3$.
The neutron detectors using $^3$He work well for both thermal neutrons and fast neutrons, but the development of alternative technologies is urgent due to shortage of $^3$He \cite{Shortage_Helium3}.
While neutron detectors using BF$_3$ can be an alternative solution, BF$_3$ works well only for thermal neutrons and has a low response rate for fast neutrons; therefore, BF$_3$ neutron detectors cannot completely replace $^3$He neutron detectors.

Gadolinium (Gd) is a well-known rare-earth element that has the highest thermal neutron capture cross-section.
The cross section of Gd for thermal neutron is around $\sim$255,000 barns (b), which is an order of magnitude larger than those of other elements, such as $\sim$5330 b for $^3$He, $\sim$940 b for $^6$Li and $\sim$3840 b for $^{10}$B \cite{Gd Thermal neutron cross-section}.
When a neutron is absorbed by a Gd atom, the excited state usually emits gamma rays with energies on the order of 8-9 MeV.
An internal conversion (IC) electron is also emitted from the de-excitation of the Gd$^*$-atom. These IC electrons have energies between 29 and 246 keV.
As these electrons and gammas can be directly detected by using a typical gaseous detector, one can design a neutron detector using this property of Gd\cite{Pfeiffer}.

A gas electron multiplier (GEM), one of the micro-pattern electron amplification detectors, was developed at CERN in the 1990s \cite{FirstIntroductionGEM}.
As the technology for the mass production of GEM foils improved, its applications in various fields also increased.
The GEM foil is manufactured by etching micrometer-sized holes into flexible copper-clad laminates (FCCL), a core component of flexible printed circuits boards (FPCBs) that are widely used in the electronics industry. 
The FCCL foil has a layer of resistive polyimide (Kapton) coated with layers of copper on both sides.
A typical GEM foil is made of this FCCL foil with a hole diameter of 70 \textmu m with 140-\textmu m pitches\cite{FirstIntroductionGEM}.
When a high voltage difference is applied between the copper layers of the GEM, a strong electric field is generated within the holes.
When electrons go near the holes, they are accelerated by the strong electric fields.
In the presence of ionizing gases, the accelerated electrons can create secondary ionized electrons, which are also accelerated within the holes, thereby creating an avalanche of electrons \cite{Standard GEM Principle, UOSsamuli}.
Typical electron amplification for a single GEM is about 20.
With a combination of three or more GEM layers, tens of thousands of gain can be achieved.

A standard GEM detector has been around in the detector community since CERN produced the 10 cm $\times$ 10 cm GEM foils in large quantities.
Recently, a Korean company successfully developed a technology for the mass production of GEM foils and began supplying them to the CMS Collaboration at CERN as part of the muon upgrade project \cite{CMSTDR}.
This company also provides customized GEM foils on demand for various applications.
Our team has been working with the company on GEM detector R\&D for many years and has successfully developed several applications, including X-ray imaging and muon detectors.

In this paper, we present our first result on neutron detection with the Gd-GEM detector that we have developed.
We'll discuss the advantage of the Gd-GEM detector as a possible thermal neutron detector.

\begin{centering}
\section{Detector Configurations}
\end{centering}
When a Gd neutron-capture occurs, an electron or gammas are emitted.
The gammas are mostly from the continuum emission peaking at around 9 MeV \cite{GdNCapture_gamma_emission}.
A few higher energy gammas deposit only fractional energy in the drift region and appear as background signals like cosmic rays.
The main signal electrons of interest are from internal conversion, with a quarter of these electrons having the specific energies of 29 keV and 39 keV \cite{IC electron from Gd neutron captrue}.
These electrons fully deposit their energy within the 10-mm drift gap due to ionization with the gas\cite{Pfeiffer}.

\begin{figure}[htbp]
    \centering
\begin{minipage}{.5\textwidth}
    \begin{tabular}{|c|c|c|c|c}
    \hline\hline
    Detector Name & Cathode Material   & Drift Gap & Additional Gd Sheet
    \\ \hline\hline
    Cu-10-GEM     & Copper             & 10 mm     & X                   \\ \hline
    Gd-Cu-10-GEM  & Copper             & 10 mm     & O                   \\ \hline
    Gd-3-GEM      & Gadolinium         & 3 mm      & X                   \\ \hline
    Gd-10-GEM     & Gadolinium         & 10 mm     & X                   \\ \hline
    \end{tabular}
\captionof{table}[Detector configurations and naming convention.]{Detector configurations and naming convention.}
\label{tab:DetectorLabelling}
\end{minipage}\hfill
\begin{minipage}{.4\textwidth}
    \includegraphics[width=0.6\textwidth]{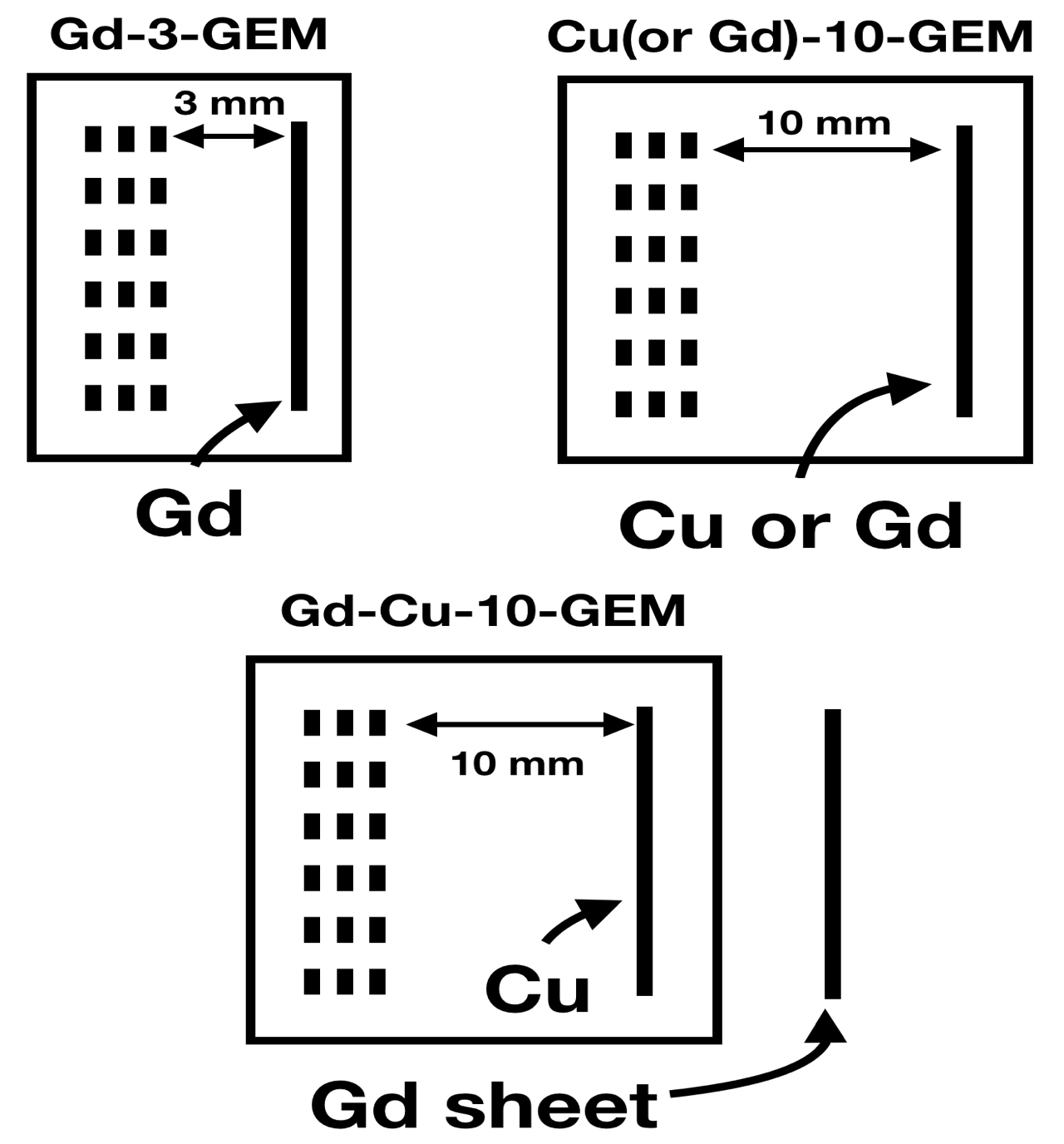}
\captionof{figure}[text for list of figures]{Schematic for each naming convention.}
\end{minipage}
\end{figure}

Four GEM chambers were assembled as shown in Table \ref{tab:DetectorLabelling}, as the notation used to identify the individual chambers, and Figure \ref{tab:DetectorLabelling}.
Cu-10-GEM and Gd-10-GEM were fabricated to compare and test the gadolinium cathode as a viable neutron converter. Gd-3-GEM and Gd-10-GEM were designed to test the effect of the drift gap size. Gd-Cu-10-GEM and Gd-10-GEM are used to identify the signal from IC electron. The active area of the gadolinium cathode is $7.8 \times 7.8 \,\rm{cm^{2}}$.

\begin{figure}[]
    \centering
      \begin{subfigure}{0.45\textwidth}
        \includegraphics[width=\textwidth]{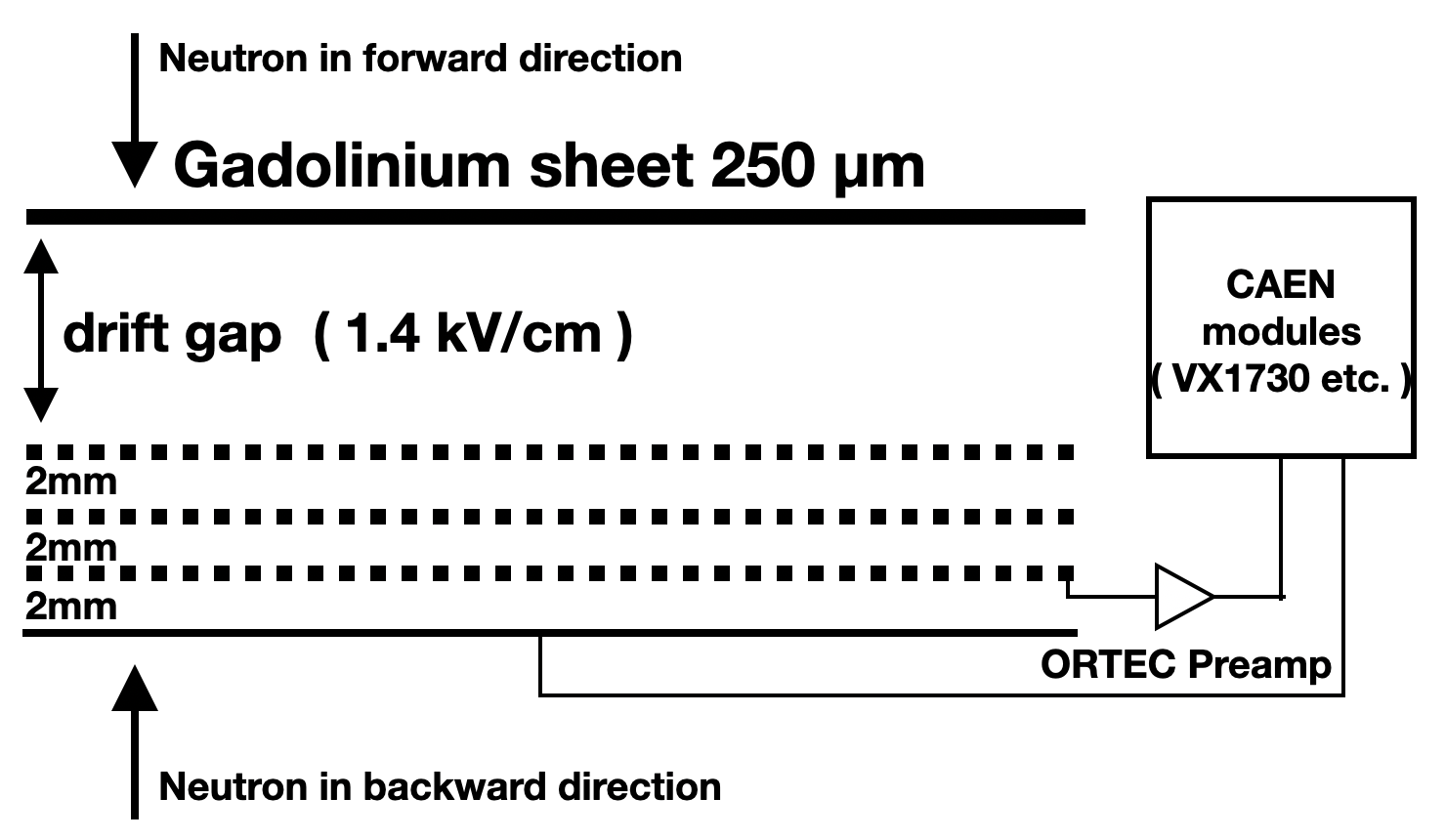}
          \caption{}
          \label{fig:GEMschematic}
      \end{subfigure}
      \hfill
      \begin{subfigure}{0.45\textwidth}
        \includegraphics[width=\textwidth]{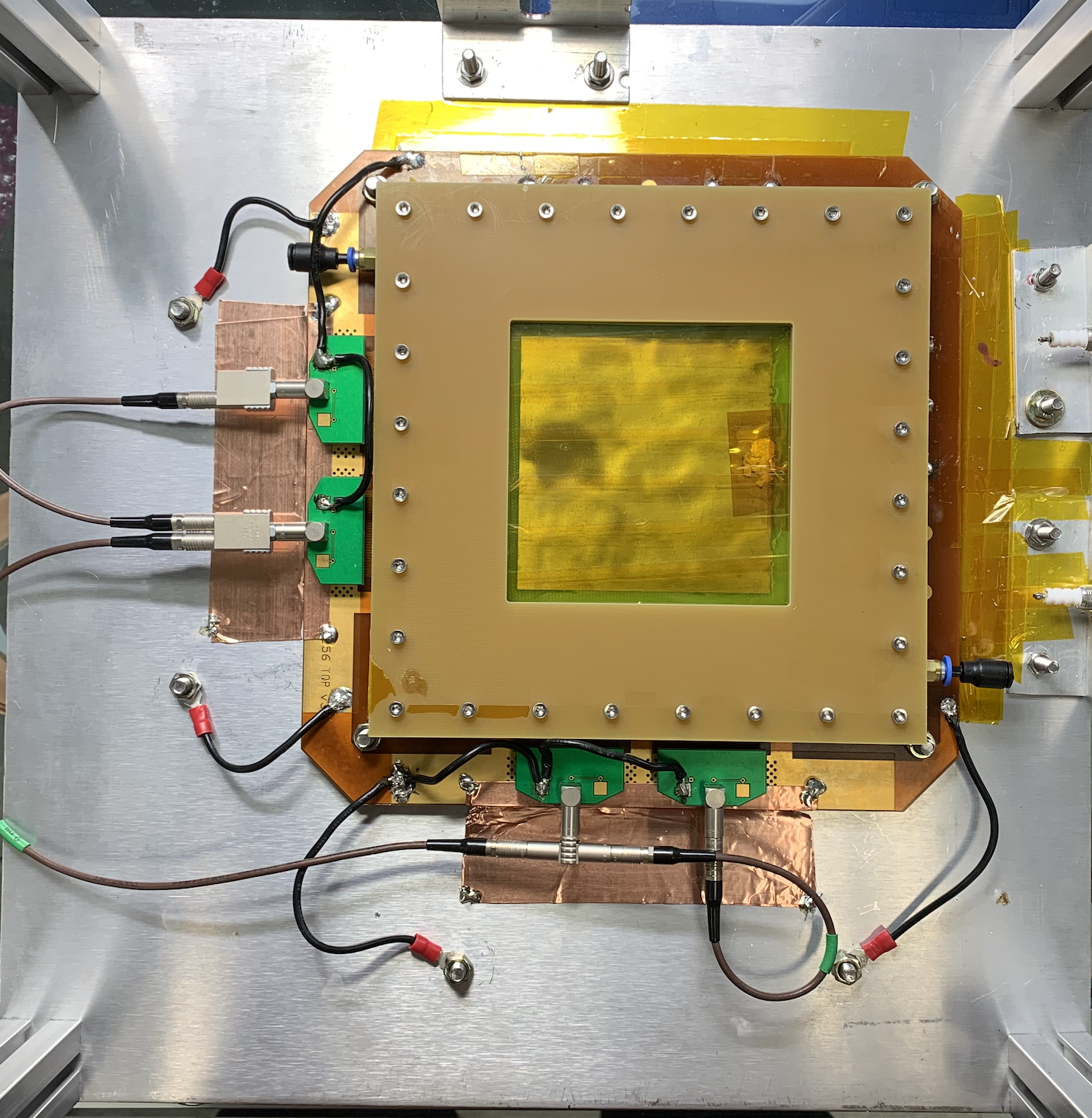}
          \caption{}
          \label{fig:GEMtopview}
      \end{subfigure}
\caption{
\label{fig:schematic}
(a) Schematic of the GEM detector with a Gd cathode and (b) photograph of the GEM detector.}
\end{figure}

A typical GEM chamber (Cu-GEM) uses a 50-\textmu m polyimide foil coated with 5-\textmu m of copper as the cathode.
Gd-GEM has a 250-\textmu m Gd sheet as a cathode layer, instead of typical Cu, for neutron capture as shown in Figure \ref{fig:GEMschematic}.

Cu-GEM and Gd-GEM were exposed to the same radioactive source under the same conditions, and a 16-bit 500-MHz fast-ADC module (CAEN VX1730) was used to obtain the distribution of integrated ADC values.

\begin{figure}[]
    \centering
      \begin{subfigure}{0.45\textwidth}
        \includegraphics[width=\textwidth]{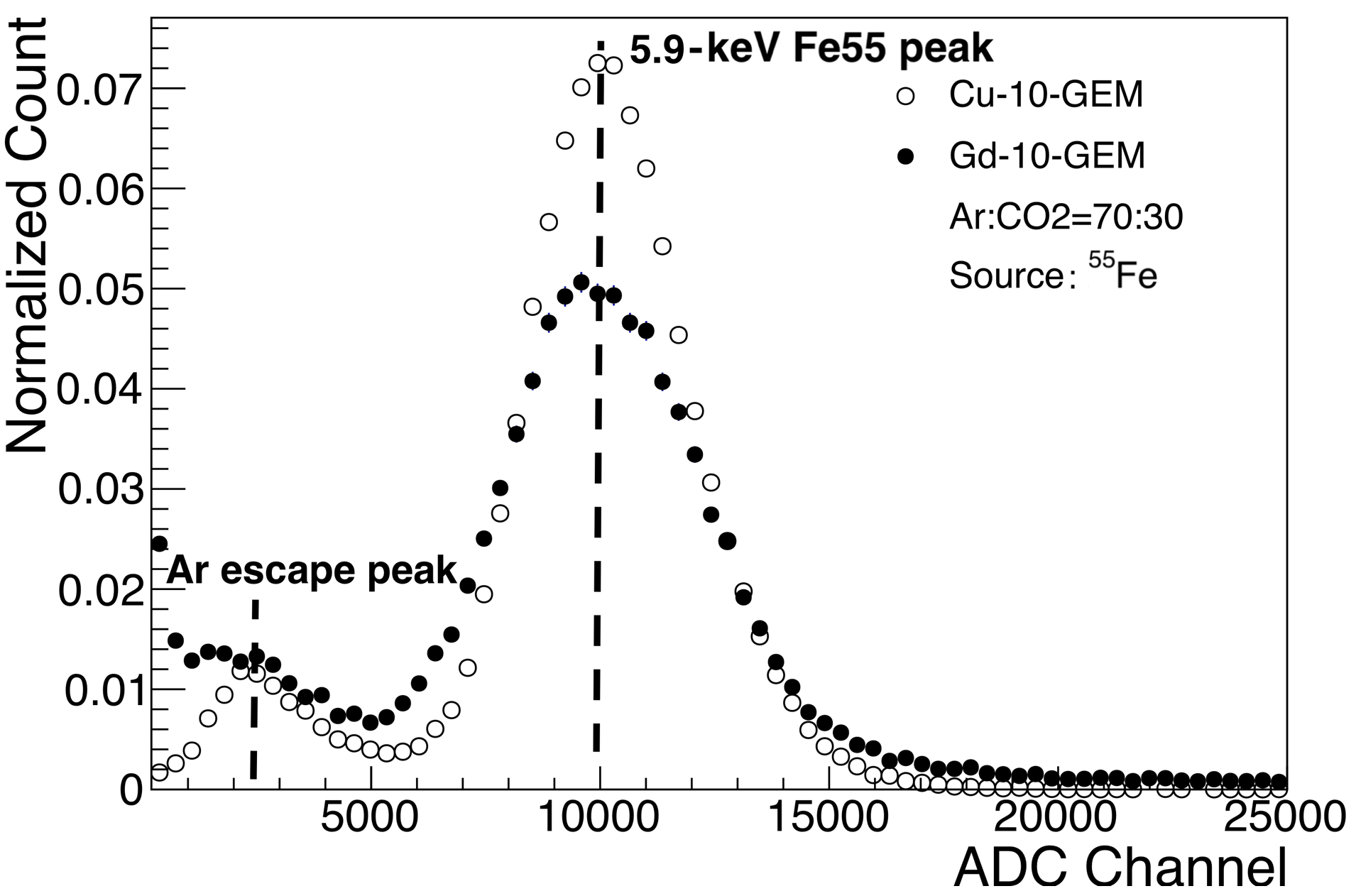}
          \caption{}
          \label{fig:Fe55Spectrum}
      \end{subfigure}
      \hfill
      \begin{subfigure}{0.45\textwidth}
        \includegraphics[width=\textwidth]{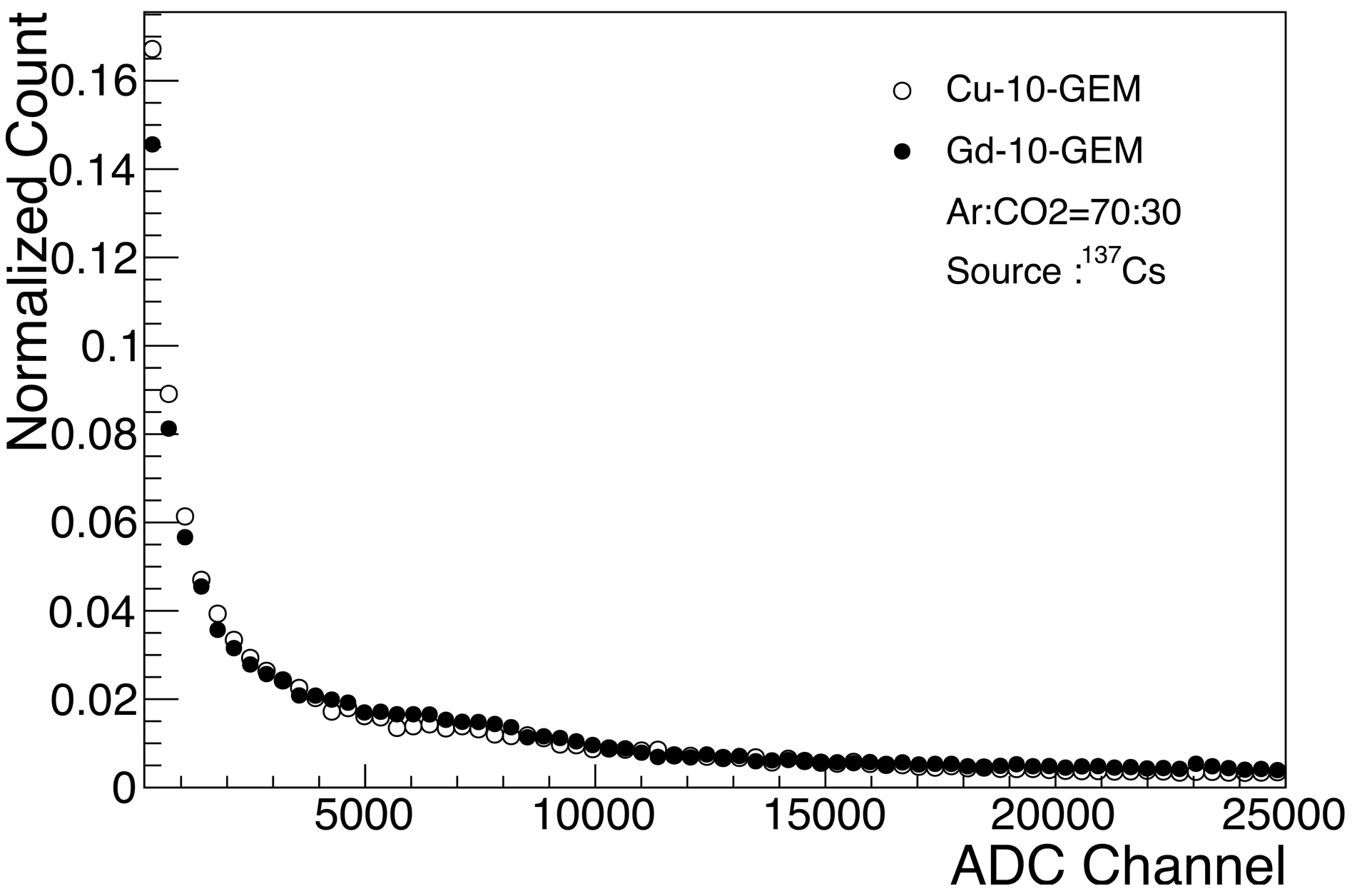}
          \caption{}
          \label{fig:Cs137Spectrum}
      \end{subfigure}
\caption{
\label{fig:SourceSpectrum}
(a) $^{55}$Fe spectra and (b) $^{137}$Cs spectra from Cu-10-GEM and Gd-10-GEM.}
\end{figure}

The spectra of Cu-10-GEM and Gd-10-GEM in the presence of a $^{55}$Fe source are shown in Figure \ref{fig:Fe55Spectrum}.
Both detectors measured nearly identical spectra, clearly showing the $^{55}$Fe peak of 5.9 keV.
However, the Gd-10-GEM detector shows a lower peak height due to the much larger (250 \textmu m) thickness of the Gd cathode compared to the Cu cathode (55 \textmu m)
The effective gains are obtained as about $6\times{10}^{4}$ by using the ionization energy of gas (70\% of Ar and 30\% of CO$_{2}$) and a radioactive source $^{55}$Fe source. 
In the Gd-10-GEM detector, the argon escape peak is partially covered due to the pedestal noise.

Figure \ref{fig:Cs137Spectrum} shows the spectra measured using a $^{137}$Cs source that emits 0.6-MeV gammas. 
Due to the small cross section at such high energy, the gamma is able to pass through gas without depositing any energy, leaving to almost identical spectra for both detectors.
Gammas with energies of a few MeVs are emitted by the AmBe source at KRISS, would also produce similar spectra.

\begin{centering}
\section{Detector Response Simulation}
\end{centering}
The simulation was performed with GEANT4~\cite{GEANT4} to test the sensitivities of the GEM detectors to neutrons. 
The GEANT4 simulation uses the FTFP\_BERT\_HP physics list and G4NDL4.6~\cite{G4NDL} and G4PhotonEvaporation5 \cite{Geant4 Photon Evaporation} data files to generate accurate simulations of thermal neutrons.
The simulation was performed over a range of Gd sheet thicknesses to test the sensitivity to neutrons as a function of thickness.
The detector was modeled as a composition of multi-layered materials as described in Table \ref{tab:modelTable}.

\begin{table}[]
\begin{tabular}{c|c|c}
  \hline\hline
    Layer   & Material          & Thickness                     \\ \hline\hline
    Cathode & Gd                & 1 \textmu m $\sim$ 5 mm          \\ \hline
    Drift Gap & Ar/CO$_{2}$(70\%/30\%) & 10 mm                         \\ \hline
    GEM     & Cu/Polyimide/Cu      & 5 \textmu m/50 \textmu m/5 \textmu m      \\ \hline
    Gas Gap & Ar/CO$_{2}$(70\%/30\%) & 2 mm                           \\ \hline
    GEM     & Cu/Polyimide/Cu      & 5 \textmu m/50 \textmu m/5 \textmu m      \\ \hline
    Gas Gap & Ar/CO$_{2}$(70\%/30\%) & 2 mm                           \\ \hline
    GEM     & Cu/Polyimide/Cu      & 5 \textmu m/50 \textmu m/5 \textmu m      \\ \hline
    Gas Gap & Ar/CO$_{2}$(70\%/30\%) & 2 mm                           \\ \hline
    Readout & Cu/FR4/Cu         & 35 \textmu m/3.2 mm/35 \textmu m       \\ \hline
\end{tabular}
\caption{Layer structure of the detector from top to bottom.} 
\label{tab:modelTable}
\end{table}

\begin{figure}[h]
\begin{subfigure}{.46\textwidth}
  \centering
  \includegraphics[width=.95\linewidth]{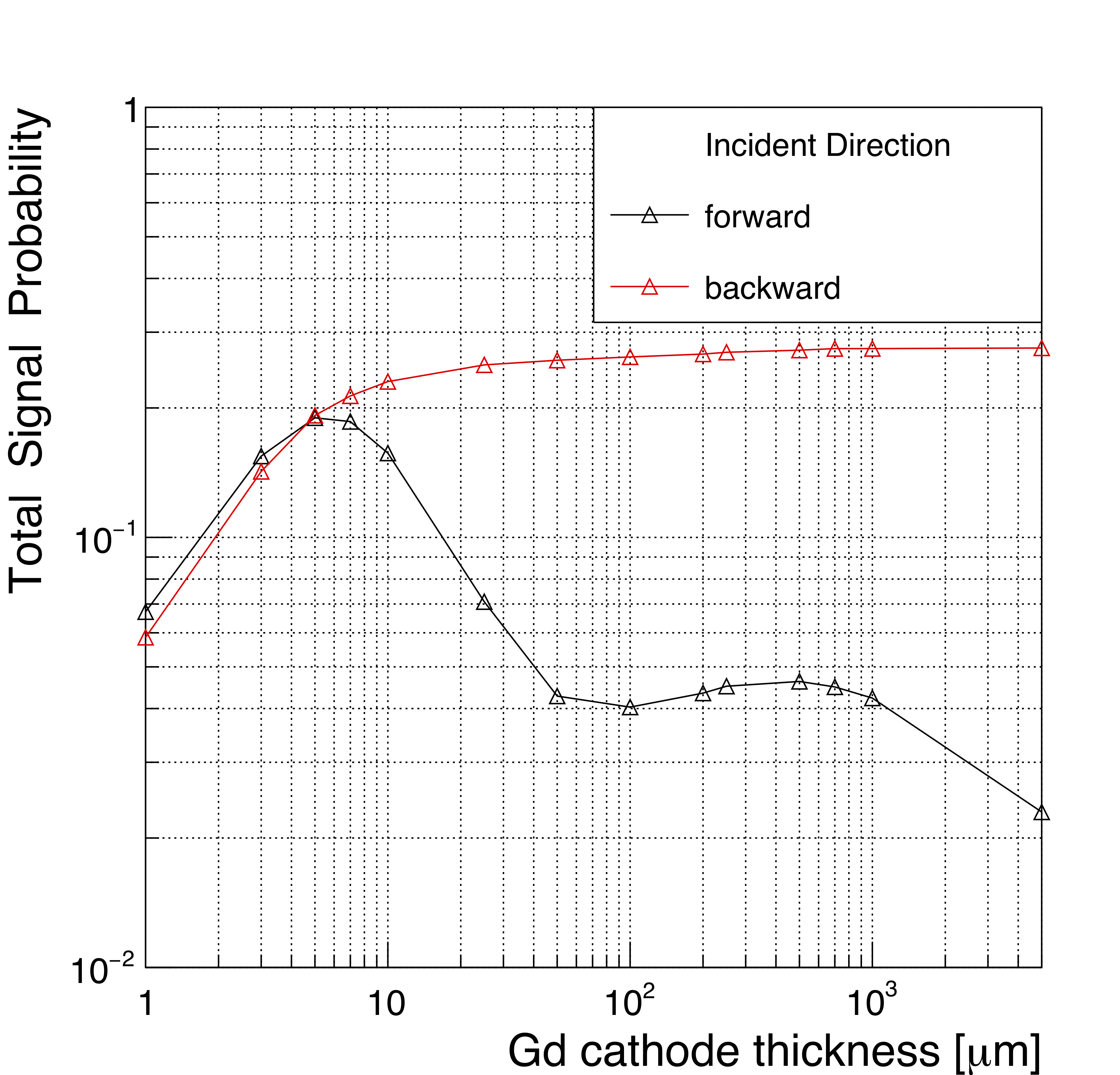}
  \caption{}
  \label{fig:gdSens}
\end{subfigure}
\begin{subfigure}{.46\textwidth}
  \centering
  \includegraphics[width=.95\linewidth]{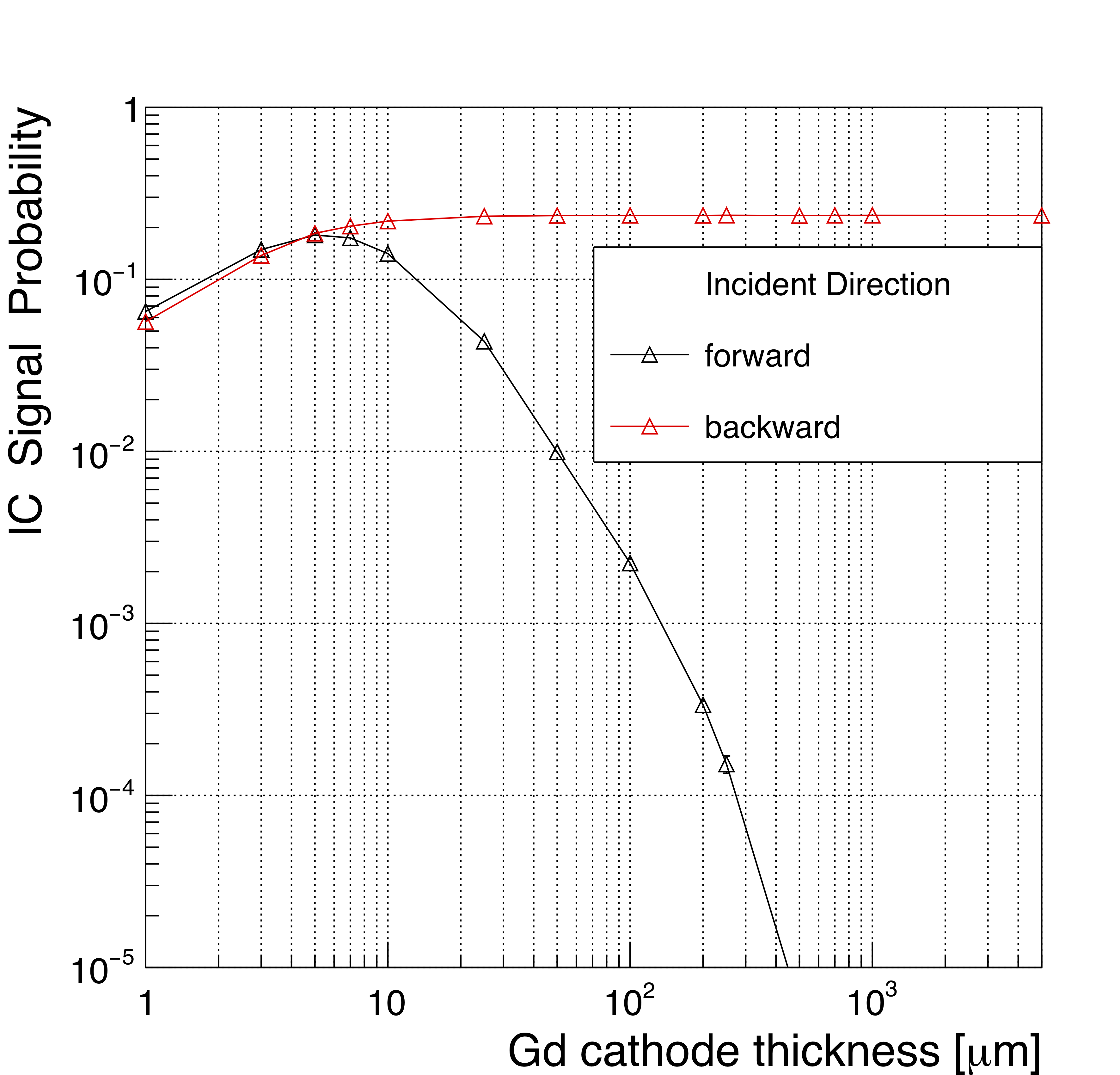}
  \caption{}
  \label{fig:gdSig}
\end{subfigure}
\caption{Probabilities that an incident neutron with an energy of 25 meV makes hits in the detector: (a) total and (b) from internal conversion only. }
\end{figure}

A particle gun was used to generate 25-meV neutrons, which are only available at the KRISS, perpendicular to the front and the backside of the detector.
The secondary electrons produced by the capture of neutrons that enter the drift gap are able to generate signals on the detector. 
The probabilities of neutrons that are able to create signals as a function of Gd thickness is shown in Figure \ref{fig:gdSens}.
The simulation was done for two different beam configurations, one for forward and the other for backward relative to the gadolinium cathode.
Figure \ref{fig:gdSig} shows the probabilities of signals created from IC electrons.
Both plots show that above a Gd thickness of 5 mm, the sensitivity decreases for neutrons in the forward direction.
Most of the electrons in the backward configuration are shown to be created by IC while for the forward configuration, the electrons are less likely to be from IC as the thickness increases. 

\begin{figure}[h]
    \centering
      \begin{subfigure}{0.45\textwidth}
        \includegraphics[width=\textwidth]{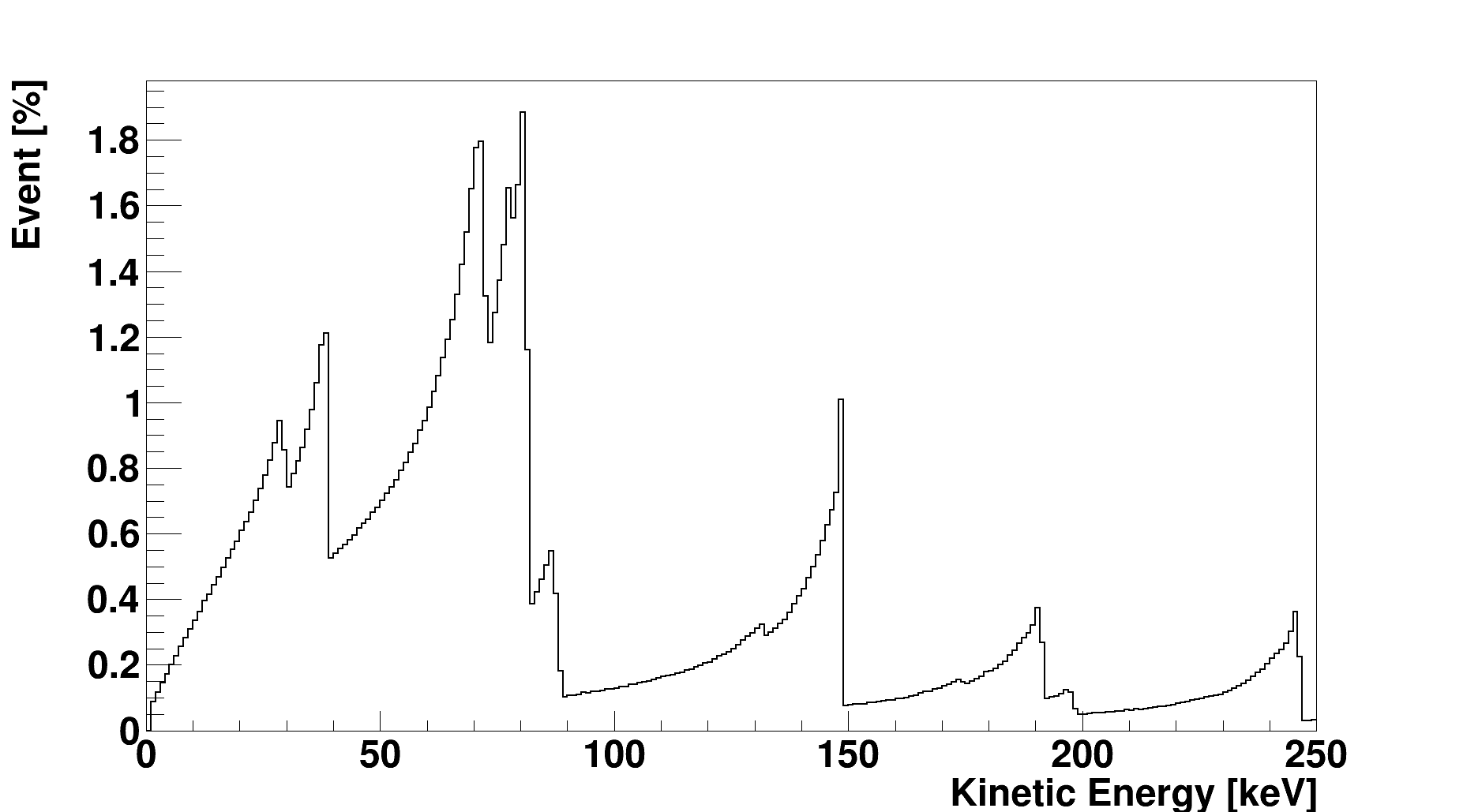}
          \caption{Kinetic energy of IC electrons entering the drift gap through Gd}
          \label{fig:Sim_ICEnergy}
      \end{subfigure}
      \hfill
      \begin{subfigure}{0.45\textwidth}
        \includegraphics[width=\textwidth]{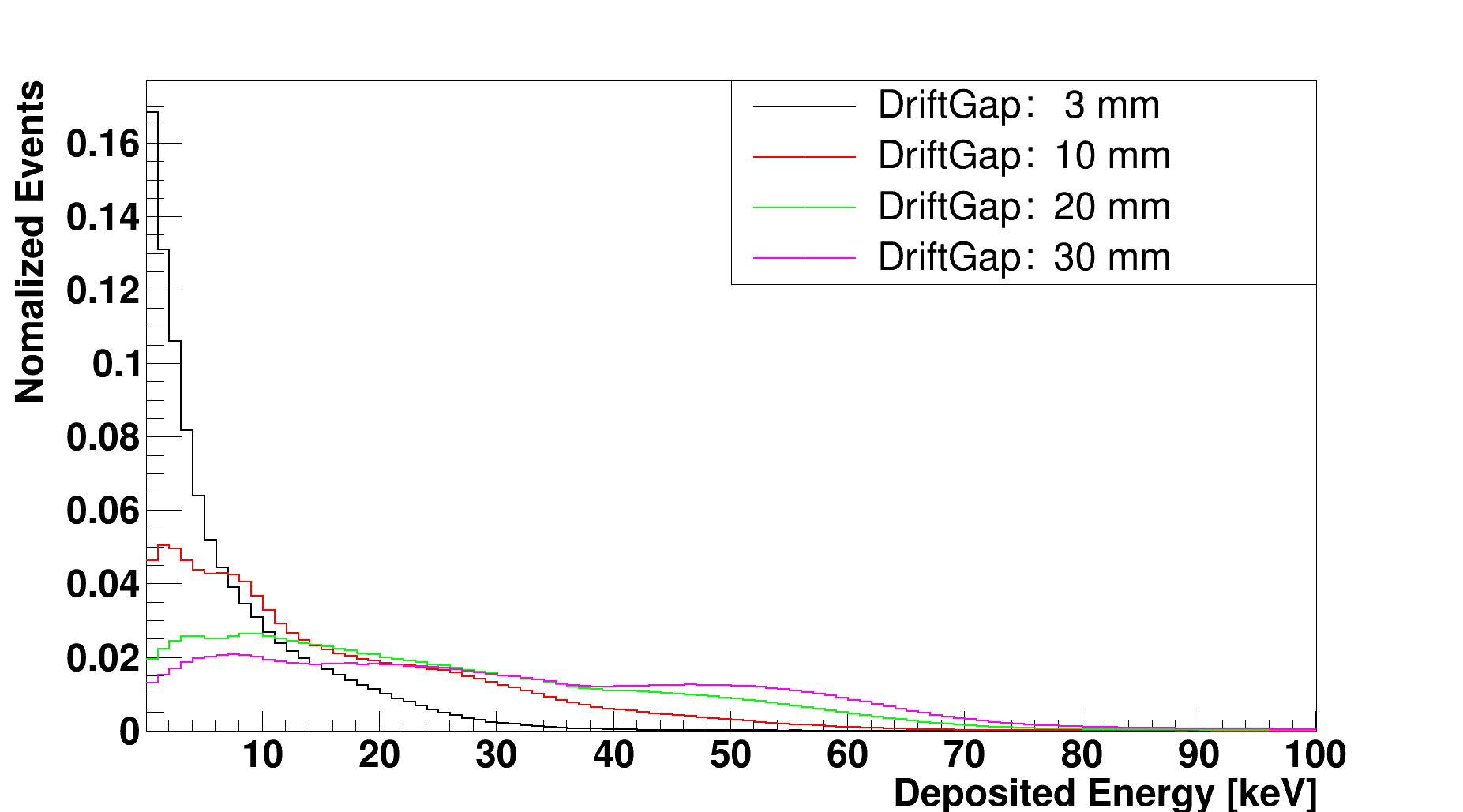}
          \caption{Energy deposited for various drift gap lengths.}
          \label{fig:Sim_EnergyDeposit}
      \end{subfigure}
\caption{
\label{fig:ICelectronAndEdep}%
Kinetic energy of IC electron and energy deposited in drift. }
\end{figure}

For a Gd sheet of 250 \textmu m in thickness, the energies of the IC electrons entering the drift are shown in Figure \ref{fig:Sim_ICEnergy}.
This highlights the discrete peaks for the IC electrons produced by the neutron capture process.
IC electrons that have energies below 40 keV have a maximum penetration depth of about 1.1 cm  \cite{Pfeiffer}.
The deposited energy is the energy lost by the particles as they traverse through the drift gap, and it is mainly due to the ionization of the gas.
This energy deposited within the drift gap is directly proportional to the readout signal current (assuming as uniform gain) \cite{Leo detector book}.
As mentioned in the previous section, our detector aims to detect the signal produced by the IC electrons emitted after neutron capture.
The IC electron energy spectrum was used to simulate the amount of energy deposited for various drift-gap lengths, and the results are shown in Figure \ref{fig:Sim_EnergyDeposit}.
Figure shows that a 3-mm gap is not sufficiently large to fully contain these electrons; thus, much larger drift gap would be preferable.
The larger drift gap, the more likely the impurities are to flow into drift gap, so the probability for sparks occurring in the chamber increases as the drift gap gets larger.
If impurities like dust burn out at the hole in the GEM foil, the damage at the hole may cause a spark or permanent defect on the GEM foil. To reduce damage to the chamber while keeping the drift gap large enough to contain the IC electrons, we selected a 10-mm drift gap.

\begin{figure}[h]
    \centering
    \includegraphics[width=0.8\textwidth]{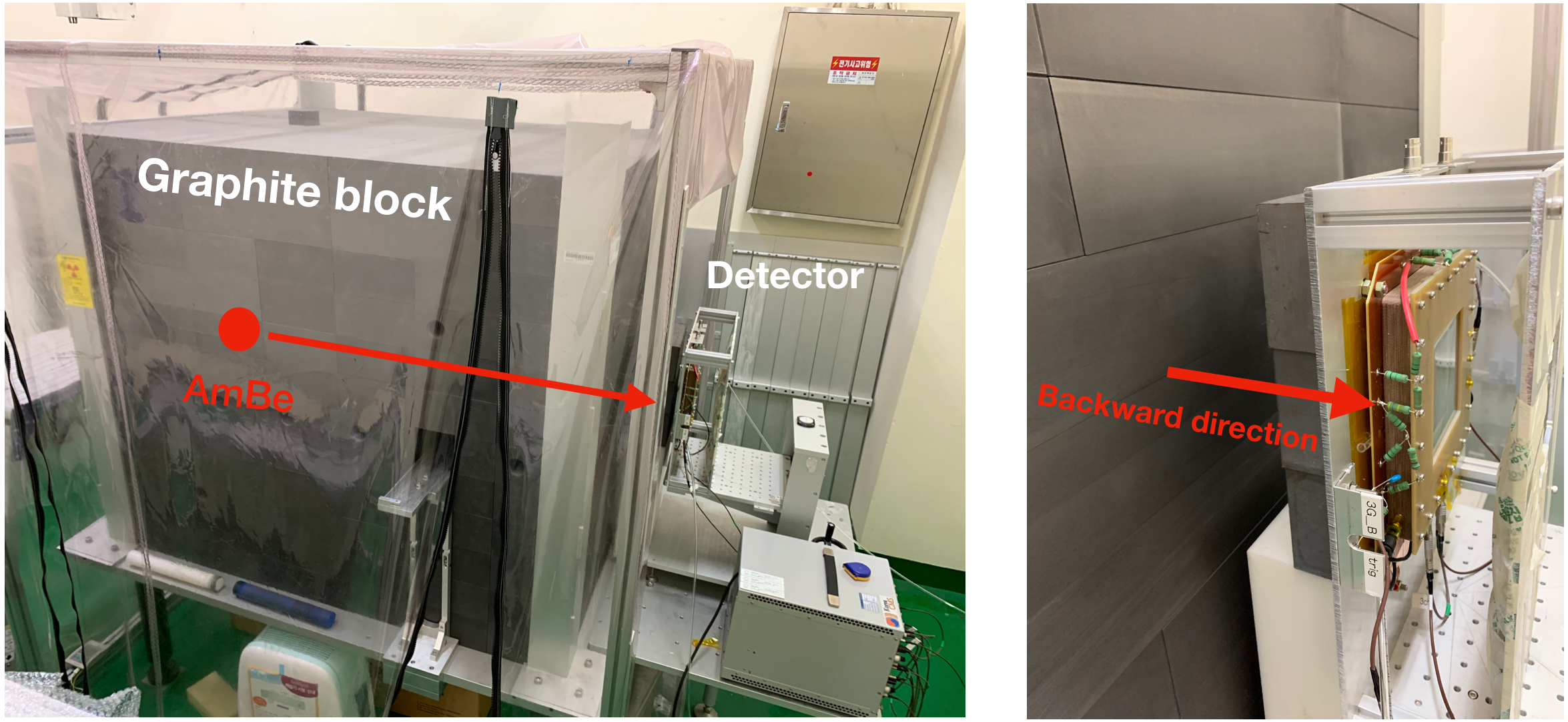}    
    \caption{Thermal neutron facility at KRISS and position of the detector.}
    \label{fig:KRISSSetting}
\end{figure}

\begin{centering}
\section{TEST BEAM RESULTS}
\end{centering}
The thermal neutron source at the Korean Research Institute of Standard and Science (KRISS) is generated from AmBe surrounded by graphite blocks.
Americium is an alpha source, and the alpha particles from the source interact with the beryllium to produce neutrons.
The neutrons from the AmBe source are fast neutrons with energies of several MeV.
At the KRISS, this AmBe source is placed in the middle of a pile of graphite blocks to lower the energy of the neutrons, as shown in Figure \ref{fig:KRISSSetting} \cite{Thermal Neutron in KRISS}.
An independent measurement (by KRISS using its neutron flux meter) of the neutron flux at the location of our GEM detector was $67.7 \pm 4.1 \rm{{n}_{th}}/{\rm{cm}}^{2}\cdot \rm{s}$ ($\rm{{n}_{th}}$ = 25-meV thermal neutron).
When we consider the area of the Gd sheet (60  ${\rm{cm}}^{2}$), about 4,000 $\rm{{n}_{th}/s}$ of thermal neutrons will traverse the Gd sheet.
High-energy gammas are also emitted by the AmBe source.
The Cu-GEM chamber will only detect the signals made by high-energy gamma emitted by the AmBe source while the Gd-GEM will detect both the continuum gammas and IC electrons emitted due to Gd-Neutron capture.
The Gd-GEM detector was mounted in two directions, one with the Gd-cathode side facing the source and the other with the Gd-cathode side facing in the opposite direction, forward and backward, respectively, as shown in Figure \ref{fig:GEMschematic}.
The experiment was repeated with the Gd-10-GEM, Gd-3-GEM, Gd-Cu-10-GEM and Cu-10-GEM in the same mounting position for both neutron irradiation directions.

\begin{figure}[h]
    \centering
    \includegraphics[width=0.8\textwidth]{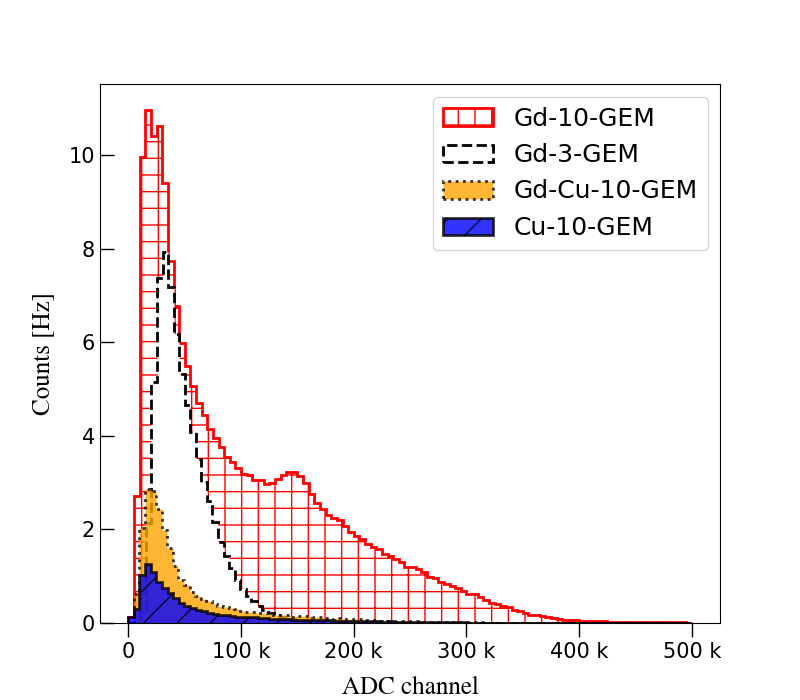}
    \caption{ADC distribution rates taken in the backward direction.}
    \label{fig:cathodetime}
\end{figure}

\begin{figure}[h]
    \centering
    \includegraphics[width=0.8\textwidth]{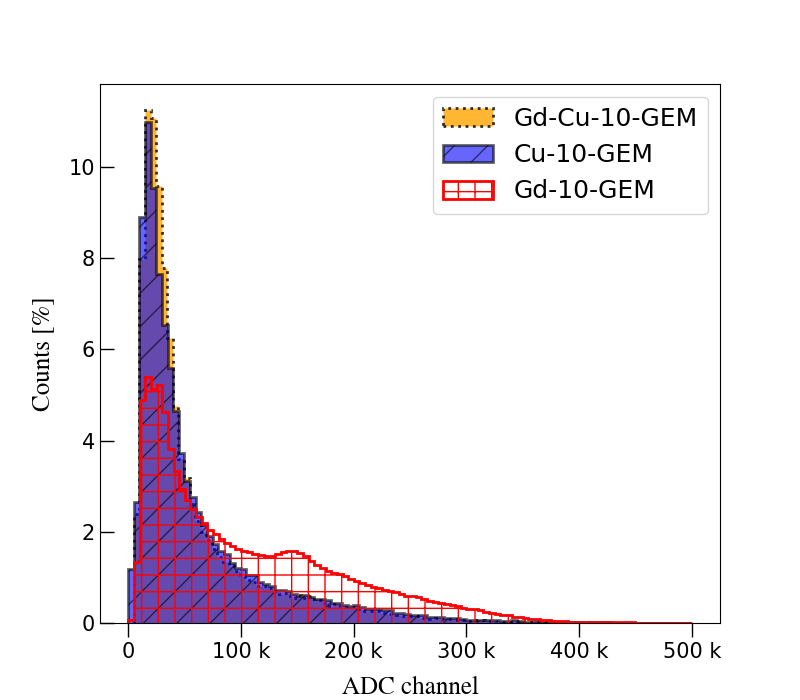}
    \caption{Normalized ADC distribution taken in the backward direction.}
    \label{fig:cathodenorm}
\end{figure}

Figure \ref{fig:cathodetime} shows the ADC distribution rates for the Gd-10-GEM, Gd-3-GEM, Gd-Cu-10-GEM and Cu-10-GEM detectors facing the backward directions.
The histograms in the Figure \ref{fig:cathodenorm} were drawn using the data in Figure \ref{fig:cathodetime}, but the data were normalized to unity.
Among the GEM chambers with various configurations, only the Gd-GEM with a 10-mm drift gap showed a small peak at around 160 k in the ADC distribution for the backward neutron irradiation direction.
This peak can be assumed to be made by the IC electrons.
In order to verify this assumption we placed, a Gd plate (10 cm $\times$ 10 cm square with 2-mm thickness) on top of the Cu-10-GEM detector (Gd-Cu-10-GEM), which will only let the gammas from neutron capture pass through to the sensitive area of the detector.
In Figure \ref{fig:cathodetime}, the Gd-Cu-10-GEM detector shows a higher rate than the Cu-10-GEM detector, but a lower rate than the Gd-10-GEM, indicating that while the Gd-10-Cu-GEM detector is also sensitive to thermal neutrons, the IC peak is not present as the IC electrons cannot penetrate to the sensitive area of the detector.

In the absence of the neutron source, the Cu-10-GEM and the Gd-10-GEM detectors measured the same event rate of 2 Hz.
In the presence of the neutron source, the total signal rates for all the detector configurations are shown in Table \ref{tab:ratetable}, and all detectors are seem to have increased signal rates.
The Cu-10-GEM detector is only sensitive to high-energy gammas directly from the AmBe source while the Gd-10-GEM detector is sensitive to both gammas and electrons from gadolinium-neutron capture, as well as the high-energy gammas from the AmBe source.
The Gd-10-GEM detector has an event rate of about 6 to 17 times higher than that of the Cu-10-GEM detector in the forward and the backward directions, respectively, clearly demonstrating the sensitivity of the Gd-10-GEM to thermal neutrons.
The total signal rate from IC electrons or gammas after the gadolinium-neutron capture process is obtained by subtracting the signal rates of the Cu-10-GEM detector from the signal rates of the Gd-10-GEM detector.
The signal rates and the corresponding statistical errors are summarized in the Table \ref{tab:ratetable}.
The active area of the gadolinium sheet is 60.84 cm$^2$, and the neutron flux is 67.7 $\pm$ 4.1$/{\rm{cm}}^{2}\cdot \rm{s}$.
The uncertainty of 6\% (=4.1/67.7) from the KRISS measurement is our main systematical error.
The statistical error that we obtained (0.0075) is an order of magnitude smaller than the systematic error (0.0604).
Our final efficiency is estimated to be $4.630 \pm 0.034(stat.) \pm 0.279(syst.) \%$.

\begin{table}[h]
    \begin{tabular}{c|c|c|c|c|c|c|c}
    \hline\hline
    Detector Configuration &
      \multicolumn{2}{c|}{\begin{tabular}[c]{@{}c@{}}Cu-10-GEM\end{tabular}} & Gd-Cu-10-GEM &
      \multicolumn{2}{c|}{\begin{tabular}[c]{@{}c@{}}Gd-10-GEM\end{tabular}} &
      \multicolumn{2}{c}{\begin{tabular}[c]{@{}c@{}}Gd-3-GEM\end{tabular}} \\ \hline
    Neutron direction  & Forward & Backward & Backward & Forward & Backward & Forward & Backward \\ \hline
    Total signal rate $[$Hz$]$ & 13 $\pm$ 0.06 & 12 $\pm$ 0.09 & 26 $\pm$ 0.07  & 80 $\pm$ 0.10 & 203 $\pm$ 0.20 & 30 $\pm$ 0.29  & 72$\pm$ 0.10 \\ \hline\hline
    \end{tabular}
\caption{Total signal rate of all detector configurations from the KRISS thermal neutron source.} 
\label{tab:ratetable}
\end{table}

In Table \ref{tab:ratetable}, the signal rate measured using the Cu-10-GEM detector showed little difference between the forward and the backward neutron irradiation directions.
The Gd-10-GEM detector measured double the signal rate in the backward irradiation direction than it did in the forward irradiation direction.
This result is qualitatively consistent with the simulation result in Figure \ref{fig:gdSens}, as the neutrons in the backward direction are much effectively detected.
When neutrons enter from the forward direction, the IC electrons have to penetrate the gadolinium cathode in order to enter the drift region.
Meanwhile, if the neutrons enter from the backward direction, the IC electrons enter the drift region directly.
Therefore, the possibility of detecting the IC electrons is higher for the backward neutron irradiation direction.

Finally, the two gadolinium-cathode GEM detectors, Gd-10-GEM and Gd-3-GEM, were compared to test the effect of the length of the drift gap.
In Table \ref{tab:ratetable} and Figure \ref{fig:cathodetime} for the backward configuration, the Gd-3-GEM detector ($72 \pm 0.10$) measured a rate about 6 times higher than the Cu-10-GEM detector ($12 \pm 0.09$) did and about 3 times lower than the Gd-10-GEM ($203 \pm 0.20$) did. 
This illustrates that the Gd-3-GEM detector is clearly sensitive to thermal neutrons, but due the shorter drift gap, the IC peak disappears as those electrons where not able to complete ionize within the detector chamber, consistent with the simulation results in Figure \ref{fig:Sim_EnergyDeposit}.

\begin{centering}
\section{Conclusion}
\end{centering}
A GEM detector was converted into a neutron detector by replacing the cathode material with gadolinium.
The thermal neutron absorption by the Gd cathode was simulated using GEANT4 software to obtain the energy distribution of internal conversion electrons and to understand the sensitivity for various detector configurations. 
Various detector configurations were developed to test and confirm the sensitivities to neutrons.
A series of beam tests on the Gd-GEM detector was performed using the standard thermal neutral sources provided by KRISS.
The neutron detection efficiency of our Gd-GEM detector was found to be around $4.630 \pm 0.034(stat.) \pm 0.279(syst.) \%$.
The readout of the Gd-GEM detector was originally designed to have 256 channels for each x-axis and each y-axis, but in this experiment, we merged them into a single channel as we were only interested in the total neutron detection efficiency.
In the future, the use of all these 2D strip channels could lead to a neutron imaging device.
The $^{3}$He and the BF$_{3}$ neutron sensors are rod types; thus, making 2D images is difficult unless one develops them as wire chambers.
The GEM has already shown its capability as a imaging sensor; thus, the Gd-GEM detector may be a good candidate for neutron imaging.
While Gd-GEM detectors will not completely replace the $^{3}$He and the BF$_{3}$ neutron detectors currently available in the industry, they will be of great interest especially for large-area neutron-beam monitors and for also as neutron imaging devices.

\begin{centering}
\section*{ACKNOWLEDGMENTS}
\end{centering}
This work was supported by the 2018 Research Fund of the University of Seoul (Inkyu Park). Also, this research was supported by a National Research Foundation of Korea (NRF) grant funded by the Korea Government's Ministry of Education, Science and Technology (MEST, 2017R1A2B4006644) (DongHyun Song, Sunyoung Yoo, Yechan Kang) and by the Basic Science Research Program through the National Research Foundation of Korea (NRF) funded by the Ministry of Education (2018R1A6A1A06024977) (Kyungeon Choi).
We also thank Dr. Jungho Kim of KRISS for his collaboration on the test beam. 


\end{document}